\documentclass[a4paper,12pt]{article}
\pdfoutput=1 
\usepackage{jheppub} 
\usepackage[T1]{fontenc} 
\usepackage{amsfonts,amssymb,amsmath}

\newcommand{\be}{\begin{equation}}
\newcommand{\ee}{\end{equation}}
\newcommand{\bea}{\begin{eqnarray}}
\newcommand{\eea}{\end{eqnarray}}
\newcommand{\beal}{\begin{aligned}}
\newcommand{\eeal}{\end{aligned}}

\title{Black Hole Thermodynamics with Conical Defects}

\author[a]{Michael Appels}
\author[a,b]{Ruth Gregory}
\author[b]{David Kubiz\v n\'ak}

\affiliation[a]{Centre for Particle Theory, Durham University,
South Road, Durham, DH1 3LE, UK}
\affiliation[b]{Perimeter Institute, 31 Caroline Street North, Waterloo, 
ON, N2L 2Y5, Canada}

\emailAdd{michael.appels@durham.ac.uk}
\emailAdd{r.a.w.gregory@durham.ac.uk}
\emailAdd{dkubiznak@perimeterinstitute.ca}

\abstract{
Recently we have shown \cite{Appels:2016uha} how to formulate a thermodynamic 
first law for a single (charged) accelerated black hole in AdS space by fixing the conical 
deficit angles present in the spacetime. Here we show how to generalise this 
result, formulating thermodynamics for black holes with varying conical
deficits. We derive a new potential for the varying tension defects: the
{\it thermodynamic length}, both for accelerating and static black holes.
We discuss possible physical processes in which the tension of a string
ending on a black hole might vary, and also map out the thermodynamic
phase space of accelerating black holes and explore their critical phenomena.
}

\preprint{DCPT-17/03}

\begin{document}

\maketitle

\section{Introduction}

The idea that black holes can behave as thermodynamical objects,
with a finite temperature and entropy has been one of the most 
fascinating yet perplexing breakthroughs in our understanding of 
these strongly gravitating objects
\cite{Bekenstein:1973ur, Bekenstein:1974ax, Hawking:1974sw}.
Not only does this insight give consistency to thermodynamical systems
in gravity, but it also gives a concrete arena for
exploring quantum gravitational effects in a controlled fashion. 
Thermodynamical considerations also give us key insights into classical aspects
of black holes in general relativity, a nice example being the black string instability
\cite{Gregory:1993vy,Gregory:1994bj}, where the onset of a classical instability 
of black brane solutions in higher dimensional gravity occurs at the point
where the entropy of the black brane drops below that of a black hole.
More generally, it seems there is a link between thermodynamic and 
dynamical stability of black holes and branes 
\cite{Gubser:2000mm,Reall:2001ag,Hollands:2012sf}.
Perhaps most pertinent for the discussion here however, are the properties
of black holes in anti de Sitter (AdS) space, where thermal
equilibrium is straightforwardly defined  \cite{Hawking:1982dh}
and physical processes correspond via a gauge/gravity duality
to a strongly coupled dual thermal field theory \cite{Witten:1998zw}.

Given the central importance of black hole thermodynamics in theoretical
gravity, it is surprising that until recently only the thermodynamics of relatively
simply systems had been explored. 
Although our catalog of exact black hole solutions is limited (mostly) to
isolated gravitating systems, there is a class of intriguing exceptions, 
given by axisymmetric solutions to the Einstein equations 
\cite{Weyl,Charmousis:2003wm}, including axisymmetric multiple black holes 
\cite{IsraelKhan,Dowker:2001dg,Gibbons:1979nf,Costa:2000kf,Herdeiro:2009vd}, 
black holes in magnetic flux tubes (including a study of thermodynamics) 
\cite{Gibbons:2013dna,Booth:2015nwa,Astorino:2016hls}, 
and most pertinently for our discussion, the accelerating
black hole. This last example has an exact solution known as the C-metric 
\cite{Kinnersley:1970zw,Plebanski:1976gy}, corresponding
to a black hole with a conical deficit emerging from one pole, or unequal conical
deficits emerging from each pole. Since it can be shown 
that the conical deficit can be replaced by a finite width topological defect 
\cite{RGMH}, the physical interpretation of the geometry is that there is a cosmic 
string `pulling' the black hole, thus accelerating it to infinity. Although this system
does not represent an isolated black hole, it is nonetheless a legitimate gravitational
solution that should be subject to the usual laws of black hole thermodynamics.

In \cite{Appels:2016uha}, we made a preliminary step towards studying thermodynamics
of accelerating black holes by considering fixed tension deficits in the context
of slowly accelerating black holes in AdS space 
(see also \cite{Dutta:2005iy,Astorino:2016xiy,Astorino:2016ybm}).
We discovered that the first law of thermodynamics was still valid, provided one
identified the mass of the black hole appropriately normalised by the conical deficit.
The fixing of tension was physically motivated by imagining that the black holes
were accelerated by cosmic strings: topological defect solutions to some 
quantum field theory \cite{Vilenkin:1984ib} whose tension is typically quantised 
in terms of the parameters of this underlying theory, and as such can only vary in
discrete amounts (and only by having the underlying topological invariant also
vary). 

However, stepping back from this specific manifestation of the C-metric, we can 
at least pose the question of what happens if we do allow the tension to vary.
Even with physical cosmic strings replacing the conical deficit, one can still 
imagine a possible scenario in which tension could jump as the merger of
two accelerating black holes along a polar axis, initially with one conical 
deficit angle between the black holes, and a larger conical deficit emerging 
from the second black hole. Indeed, initial investigations of the interactions and capture
of cosmic strings by black holes studied precisely the thermodynamics of the
system of a black hole with a string \cite{AFV,Martinez:1990sd,Bonjour:1998rf}
(although in these cases, the black holes were not accelerating). Thus,
allowing a cosmic string to have varying tension would appear to be a
desirable feature in a full exploration of the thermodynamics of accelerating
black holes.

In this paper we formulate laws of accelerating black hole thermodynamics,
allowing for a varying tension. We perform our analysis in AdS in order to
utilise solutions that have only one horizon, although the first law
we derive will still apply to any horizon locally. To some extent, this is matter of
technical convenience; we define temperature via a process of Euclideanization 
thus if more than one horizon is present, the Euclidean section will contain
conical singularities. These Euclidean conical singularities are distinct from those
of the accelerating black holes of the C-metric -- they are not the result of 
sources, but instead should be viewed as necessary singularities arising from
analytic continuation. They reflect the fact that different horizons can be at different
temperatures (though can be integrated in a controlled fashion \cite{Gregory:2013hja}).

We first present thermodynamics of black holes
with varying conical deficit in the absence of acceleration, partly as a warm-up
exercise, but more importantly to motivate and isolate the impact of a varying
tension deficit. We show that the correct first law of thermodynamics includes
a ``$\lambda\delta\mu$'' term, where $\mu$ is the tension of the deficit, and
$\lambda$ the corresponding thermodynamic potential. We then turn to the
charged accelerating black hole and derive the general first law for accelerating 
black holes:
\be
\delta M = T \delta S + V \delta P + \Phi\delta Q - \lambda_+ \delta \mu_+
- \lambda_- \delta \mu_-
\label{firstlawabhm}
\ee
allowing not only variation of entropy and charge, but also the cosmological 
constant (encoded in the thermodynamic pressure $P$
\cite{Teitelboim:1985dp,Kastor:2009wy,Dolan:2010ha,Cvetic:2010jb,
Dolan:2011xt,Kubiznak:2012wp,Kubiznak:2016qmn})
and allowing the tensions along each axis $\mu_\pm$ to vary independently.
We derive the corresponding thermodynamic lengths for each tension,
as well as the appropriate expressions for the normalised thermodynamic mass
and electromagnetic potential. Analogous to the rotating AdS black hole 
\cite{Gibbons:2004ai}, we find that the electromagnetic potential and thermodynamic
mass are modified beyond the simple geometric conical deficit factor. 
Finally, we explore the physical consequences of our results.

\section{Thermodynamics with Conical Deficits}\label{sec:singlemu}

Before discussing the accelerating black hole, it is interesting to revisit the
thermodynamics of a black hole with a `cosmic string': a spacetime first studied
by Aryal, Ford and Vilenkin (AFV) \cite{AFV}. AFV considered a conical deficit
through a Schwarzschild black hole:
\be
ds^2 = f(r) dt^2 - \frac{dr^2}{f(r)} - r^2 d\theta^2
- r^2 \sin^2\theta \left (\frac{d\phi}{K}\right ) ^2
\label{afv}
\ee
where $f(r) = 1 -2m/r$. They considered a first law of thermodynamics to argue
that the entropy of the black hole remained at one quarter of its area, now
containing a factor of $K$: $S = \pi r_+^2/K$. The thermodynamics of a
black hole with a string was also considered in greater thoroughness by
Martinez and York \cite{Martinez:1990sd}, although the `tension' of the cosmic string,
(defined later in \eqref{afvtension}) was held fixed. The only context in 
which a `varying' tension was considered was in \cite{Bonjour:1998rf}, 
where the varying tension was produced by the capture of a moving cosmic string 
by a black hole, and it was argued that in the collision of a black hole and cosmic
string, the black hole would retain a portion of the string thus increasing its mass.

We will revisit this static system first, as a means of exploring the impact of
varying tension on black hole thermodynamics. We will consider
a charged black hole represented by the metric \eqref{afv}, with
\be
f(r) = 1-\frac{2m}{r} + \frac{e^{2}}{r^{2}} + \frac{r^{2}}{\ell^2}\;, \qquad
\text{and} \quad {B} = - \frac{e}{r} dt\;.
\ee
The parameters $m$ and $e$ are related to the black hole's mass, 
and charge respectively, $B$ is the Maxwell potential, and we allow 
for a negative cosmological constant via $\ell = \sqrt{-\Lambda/3}$.

In order to treat varying tension we leave the parameter $K$ in \eqref{afv} unspecified. 
Typically, this parameter would simply be unity (or a function of rotation in the
Kerr-AdS case), however, by keeping $K$ explicitly in the metric we can study conical 
defects through a well-behaved system in a straightforward manner. 

Examining the geometry near $\theta_+ = 0$ and $\theta_- = \pi$ reveals 
how the parameter $K$ relates to the conical defect. Near the poles, the metric becomes
\begin{align}
ds_{\mathrm{II}}^2 = r^2 \left[ d\vartheta^2 + \frac{\vartheta^2}{K^2} d\phi^2\right],
\end{align}
on surfaces of constant $t$ and $r$, where $\vartheta = \pm (\theta - \theta_\pm)$ 
is the `distance' to either pole. If $K\neq 1$, 
there will be a conical defect along the axis of revolution, 
which corresponds to a cosmic string of tension
\be
\mu = \frac{\delta}{8\pi} = \frac14 \bigg[1-\frac{1}{K}\bigg],
\label{afvtension}
\ee
where $\delta$ is the conical deficit. The interpretation of tension is justified by
analysing the equations of motion for an actual cosmic string vortex in the
presence of a black hole \cite{AGK}, where \eqref{afv} was obtained as the
asymptotic form of the metric outside the string core. 
A tensionless string corresponds to a 
regular pole, $K=1$, and in this metric, the tension along either polar axis is equal, 
allowing simultaneous regularisation of the two poles. The static black hole is
inertial, as the deficits balance each other out. This exercise provides 
insight into the role $K$ plays within a metric. Different values for this 
parameter determine the severity of an overall defect running through the black hole. 

Now let us consider the temperature and entropy of the black hole. We compute
$T$ by demanding regularity of the Euclidean section of the black hole
\cite{Gibbons:1976ue}, giving
\be
T = \frac{f'(r_+)}{4\pi} = \frac{1}{2\pi r_+^2} \left [
m - \frac{e^2}{r_+} + \frac{r_+^3}{\ell^2}\right]
\ee
thus
\be
2TS = \frac{m}{K} - \frac{e}{r_+}\left (\frac{e}{K} \right ) 
+ 2\left (\frac{3}{8\pi \ell^2} \right ) \left (\frac{4\pi}{3K} r_+^3\right ) 
= M - \Phi_H Q + 2PV
\ee
gives a Smarr formula \cite{Smarr:1972kt} for the black hole, where $M=m/K$ is
the mass of the black hole, $Q = \frac{1}{4\pi}
\int \star dB=e/K$ is the charge on the black hole, 
$\Phi_H = e/r_+$ the potential at the horizon, and $P = 3/8\pi \ell^2$, 
$V = 4\pi r_+^3/3K$ the thermodynamic pressure and volume respectively 
\cite{Teitelboim:1985dp,Kastor:2009wy,Dolan:2010ha,Cvetic:2010jb,
Dolan:2011xt,Kubiznak:2012wp,Kubiznak:2016qmn}.

Now let us consider the effect of changing the parameters of the black hole
a small amount; the location of the horizon of the black hole will also shift
so that $(f+\delta f)= 0$ at $(r_++\delta r_+) $:
\be
0= f'(r_+) \delta r_+ - \frac{2\delta m}{r_+} 
+ \frac{2e\delta e}{r_+^2}  - 2r_+^2 \frac{\delta \ell}{\ell^3}
\label{simplevary}
\ee
However, we can now replace the variation of the parameters
$m, e, \ell$ with the variation of the corresponding thermodynamic charges
$M,Q,P$, and the variation of $r_+$ with that of entropy,
with the important proviso that we must allow for the variation
of tension through $K$. Thus $\delta m = K \delta M + M \delta K$ etc.\
and $\delta K = 4K^2\delta \mu$ from \eqref{afvtension}.
After some rearrangement, \eqref{simplevary} gives our  
first law of Thermodynamics with varying tension:
\be
\delta M = T \delta S + V \delta P + \Phi_H \delta Q
-2 \lambda \delta\mu
\label{firstlawwithK}
\ee
where
\be 
\lambda = \left ( r_+ - KM  \right)
\label{TDlengthnoacc}
\ee
is a \emph{thermodynamic length} conjugate to the string tension. 
This is the central result of our paper -- that string tension (in this case
equal along each axis) can be thought of as analogous to a thermodynamic 
charge that therefore has a corresponding thermodynamic potential. 
Rather than write a single ``$\lambda\delta\mu$'' term, instead we write
two ``$\lambda\delta\mu$'' terms, referring to the deficits emerging from each
pole. Although these are obviously equal in this case, for 
accelerating black holes the string tensions from
each pole are not the same, and we therefore separate these out now.

Surprisingly perhaps, this thermodynamic length is not simply the 
geometric length $r_+$ of the string from pole to singularity. 
Instead, the mass-dependent adjustment
emphasises this is a potential, rather than just an internal energy
term that might more appropriately be placed on the left hand side
of the equation. 

To see why this is the case, consider the first law in action: the 
capture, and subsequent escape, of a cosmic string by a black hole. 
This example was first proposed in \cite{Bonjour:1998rf} in the
case of a charged vacuum black hole. 
The idea is that the string is moving and gets briefly captured by
the black hole. In the capture process, the internal energy of the black
hole should remain fixed: the physical intuition is that if a cosmic string
were to pass through a spherical shell of matter, energy conservation 
would demand that the spherical shell still have the same total energy
throughout the process,
thus either it would become denser, or its radius would increase.
Of course, in the case of the spherical shell, the cosmic string would simply
transit through, leaving the system. For the black hole however, we will
see this is not the case, and we have the interpretation of a segment of 
string having been captured by the black hole, with the black hole increasing
its mass accordingly. (This process was considered in the probe
limit in \cite{Lonsdale:1988xd,DeVilliers:1997nk}.)
We therefore consider the vacuum Reissner-Nordstrom (RN)
\be
f(r) = 1  - \frac{2m}{r} + \frac{e^2}{r^2}
\ee
with the charge of the black hole being defined via
$Q = e/K$, and the electric potential being
$\Phi = e/r_+$.

Let us suppose that the string is light, or $\mu \ll 1$, then in the first 
stage where the black hole captures the string, fixing $M$
and $Q$ implies $\delta m = 4 m \delta \mu$ and $\delta e = 4e\delta \mu$
(to first order in $\mu$). Thus
\be
T \delta S = \frac{r_+-r_-}{4\pi r_+^2} \left [
2\pi r_+ \delta r_+ - 4\pi r_+^2 \delta\mu \right ]
= (r_+-r_-) \delta \mu = 2\lambda \delta \mu
\ee
as required. Interestingly, because the internal energy has been fixed,
the event horizon has to move outwards to compensate for the conical 
deficit. Since the entropy contains just one factor of $K$, but two of $r_+$,
the net effect is an increase of entropy, indicating this is an irreversible
thermodynamic process. The one interesting exception being an extremal black
hole.

In the second step, the string pulls off the black hole, so $\delta \mu = -\mu$,
and since the string is uncharged, $\delta Q$ must remain zero, and $e$ 
returns to its original value. However, since entropy cannot decrease, 
$M$ must increase
\be
\delta M = T\delta S + 2(r_+-m)\mu
= \frac{(r_+-r_-) \delta r_+}{2r_+} + 2(r_+-r_-)\mu 
\ee
In \cite{Bonjour:1998rf}, we supposed that $m$ did not change, leading to
an increase in $M$ of $4m\mu$, that we incorrectly stated was the mass
of the string behind the event horizon (this is only true for the uncharged black 
hole). Instead, it seems more physically accurate to suppose that $r_+$ does
not decrease, as otherwise the local geodesic congruence defining the event
horizon would appear to be contracting in contradiction to the area theorem.
In this case, $\delta M = 2 (r_+-r_-) \mu$, or the length of cosmic string
trapped between the inner and outer horizons. Even if one allows the 
local horizon radius to shrink while maintaining constant entropy,
$\delta M = (r_+ - r_-)\mu$: half the former amount, but still an increase
of mass due to the capture of a length of cosmic string.

\section{Thermodynamics of the C-metric}

We would now like to turn to the more general case where the deficits
emerging from each pole are no longer equal. The disparity in the 
size of the deficit produces an overall force in the direction of the largest
conical {\it deficit}, and the geometry is described by the {\it C-metric}
\cite{Kinnersley:1970zw,Plebanski:1976gy,Dias:2002mi,Griffiths:2005qp,Podolsky:2002nk}.
Typically, C-metrics have both black hole and acceleration horizons, 
however, for simplicity we wish to restrict ourselves to having only
the black hole horizon, in order that we have a well-defined temperature
for the space-time. Such a geometry is called the {\it slowly accelerating
C-metric} \cite{Podolsky:2002nk}, and we first review this geometry before 
turning to its thermodynamics.

\subsection{The Slowly Accelerating C-metric}

In this subsection we briefly review the ``slowly accelerating'' C-metric. 
The general C-metric 
\cite{Kinnersley:1970zw,Plebanski:1976gy,Dias:2002mi,Griffiths:2005qp}
represents either one or two accelerating black holes with unequal
conical deficits extending from each pole of the black hole either to infinity or
an acceleration horizon. Although the C-metric is well-known among 
relativists, there are features of the specific form we will be using that are
worth highlighting, discussing how they depend on the parameters of the solution.

For this paper, we will neglect rotation, hence the general charged, accelerating 
AdS black hole is represented by the metric and gauge potential \cite{Griffiths:2005qp}:
\be
\beal
ds^2&=\frac{1}{\Omega^2}\Biggl\{ 
f(r) dt^2
-\frac{dr^2}{f(r)} - r^2 \left[ \frac{d\theta^2}{g(\theta)} 
+ g(\theta)\sin^2\theta \frac{d\phi^2}{K^2}\right]\Biggr\}
\; , \qquad
{B}= -\frac{e}{r} dt,
\eeal
\label{dimlessmtrc}
\ee
where
\be
\Omega=1+A r \cos\theta\;,
\ee
is the conformal factor that will determine the location of the AdS
boundary (which is {\it not} at ``$r=\infty$'') in terms of the 
$\{r,\theta\}$ coordinates, and the functions $f(r), g(\theta)$ are
\be
\beal
f(r)&=(1-A^2r^2)\Bigl(1-\frac{2 m}{r}+\frac{e^2}{r^2}\Bigr)
+\frac{r^2}{\ell^2} \;,\\
g(\theta)&=1+2mA \cos\theta+ e^2A^2 \cos^2\theta\,.
\eeal
\label{fandg}
\ee
Note we are using the Hong-Teo \cite{Hong:2003gx} style coordinates
that have a direct relation to the usual Boyer-Lindquist coordinates
for the Kerr metric.

This spacetime has potential conical singularities on the axes $\theta_+=0$
and $\theta_-=\pi$. As before, we associate these conical singularities
to cosmic strings with tensions $\mu_\pm = \delta_\pm/8\pi$, 
where $\delta_\pm$ are the conical deficit angles at each pole,
found by expanding near $\theta=\theta_\pm$:
\be
4\mu_\pm =  1 - \frac{g(\theta_\pm)}{K}
= 1 - \frac1K \left (1 \pm 2mA + e^2A^2 \right)
\label{tensions}
\ee
Clearly, $\mu_+ \leq \mu_- \leq 1/4$, and in order to avoid negative tension
defects, we require $\mu_+ \geq 0$. For the majority of the paper, 
we fix $K$ by demanding that one axis be regular ($\mu_+=0$), and the 
other to have a positive tension cosmic string, i.e.
\be
K = g(\theta_+) = 1 + 2mA + e^2A^2\quad \Rightarrow \qquad
\mu_-= \frac{mA}{K}
\ee
however in the interests of generality we will keep $\mu_+$ and $\mu_-$ as
independent variables.

In order to understand the structure of this accelerating black hole, 
it is useful to consider the effects of the various parameters. 
Although we suspect $m$ and $e$ play the roles of mass and charge
for the black hole, the original form of the C-metric has these variables
defined slightly differently. In order to check this, consider the Komar 
integral for the mass using the
method described in \cite{Magnon:1985sc}
\be
M_K = \frac{1}{8\pi} \int_{S^2} \star d\xi - \frac{1}{4\pi}
\int n_a R^a_b \xi^b \sqrt{h} d^3x = \frac{m}{K}
\label{komar}
\ee
where $\xi = \partial_t$ is a timelike Killing vector field, the first integral is performed 
over a $\{\theta,\phi\}$ surface of topology $S^2$ considered to be near infinity, 
and the second integral is performed throughout the interior of this surface at 
constant `time' (as defined by $\xi$).
Note this integral is actually independent of $A$, hence we can associate
$m$ with the bare mass of the black hole, whether accelerating or not.
Similarly, $e$ is related to the charge of the black hole:
\be
Q = \frac{1}{4\pi} \int_{S^2} \star dB = \frac{e}{K}
\label{Qdef}
\ee

\begin{figure}
\includegraphics[scale=0.3]{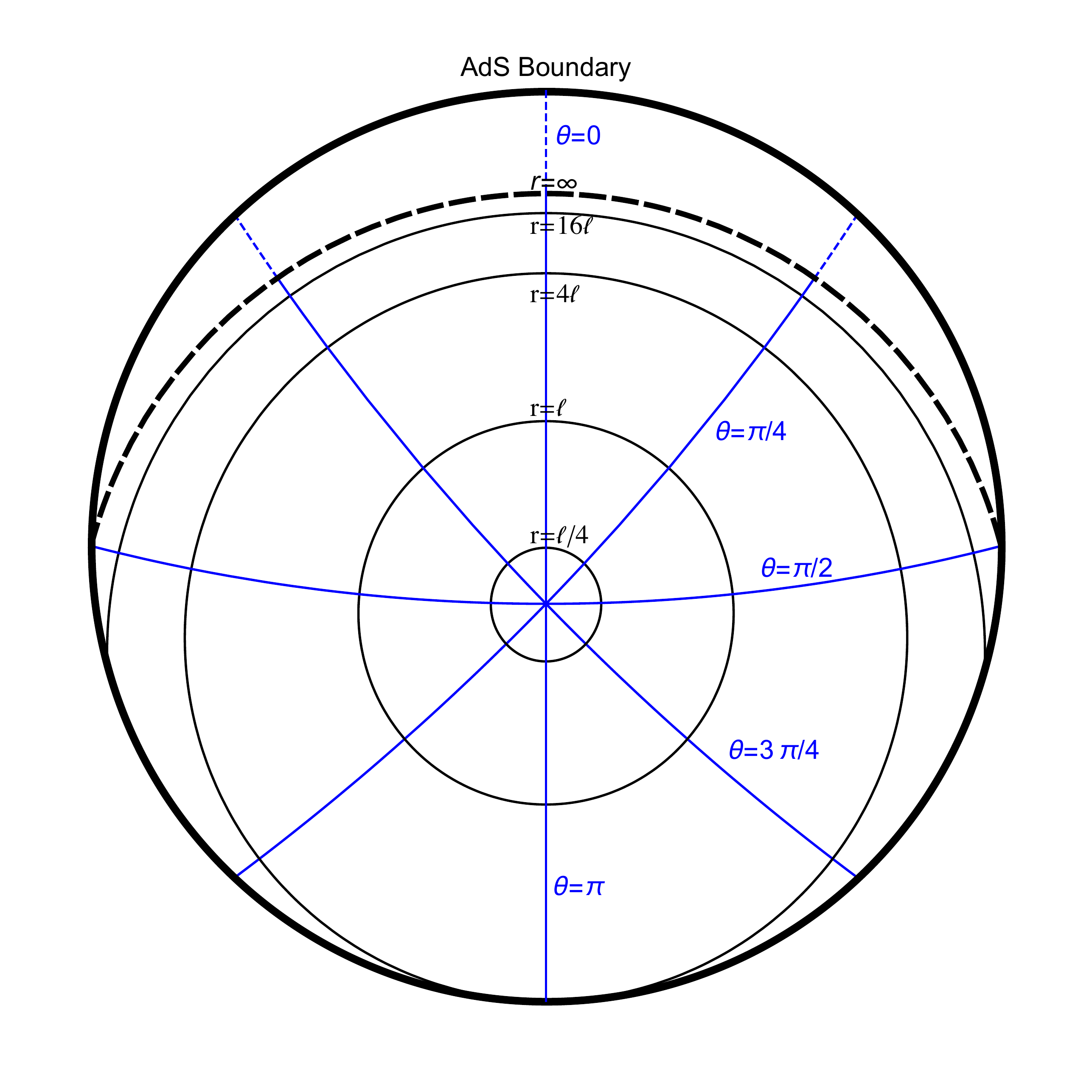}
\includegraphics[scale=0.5]{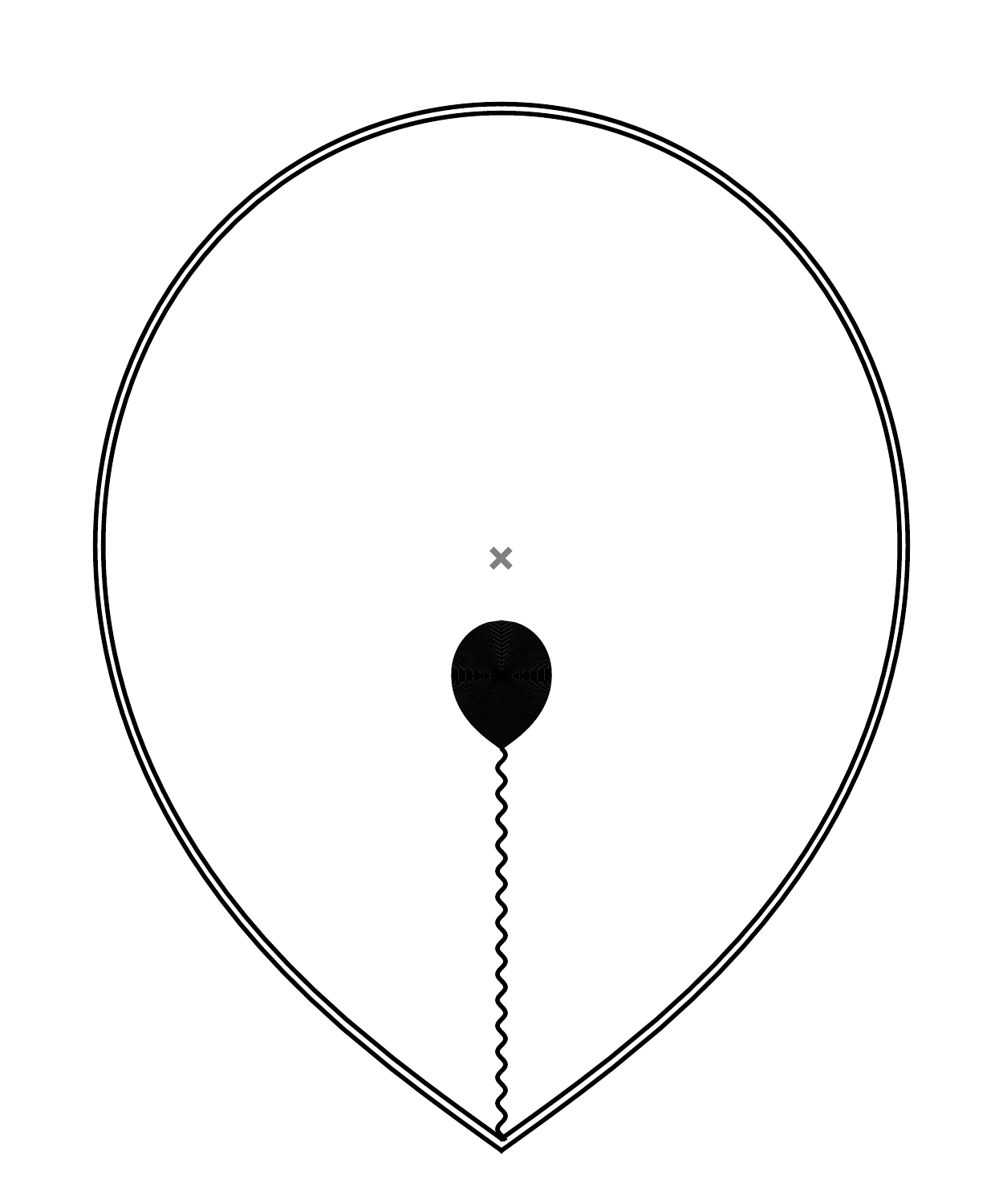}
\caption{{\bf Left}: The slowly accelerating Rindler spacetime shown here with $A\ell = 1/4$,
and $\ell=1$ for simplicity. The spatial sections of
AdS have been compactified to a Poincar\'e disc, with the constant $r$ Rindler 
coordinate indicated in black and constant $\theta$ in blue. The origin of the Rindler
coordinates is clearly visible as being displaced from the centre of the disc, with the
limit of the $r-$coordinate being the thick dashed black line.
{\bf Right}: The black hole distorts the Poincar\'e disc with a conical deficit, and is displaced 
from the origin of AdS. The spacetime is again static, and a cross section is shown.
}
\label{fig:accbh}
\end{figure}

We have already seen how the parameter $K$ is related to the conical
deficits, thus the only remaining parameter to explore is $A$. In order
to understand the structure of the spacetime and the relevance of $A$,
first consider the Rindler space found by
setting $m=e=0$ in \eqref{dimlessmtrc}:
\be
ds^2=\frac{1}{\Omega^2}\Biggl[ 
\left ( 1 + \frac{r^2}{\ell^2} (1-A^2 \ell^2) \right) dt^2
-\frac{dr^2}{1 + \frac{r^2}{\ell^2}(1-A^2\ell^2)} - r^2 \left ( d\theta^2
+ \sin^2 \theta d\phi^2\right)\Biggr]
\label{rindler}
\ee
This spacetime no longer has a conical singularity and is locally pure AdS,
however in these coordinates the boundary of AdS is not at $r=\infty$, but
at $r = -1/(A\cos\theta)$. For $\theta$ in the southern hemisphere, this occurs
at {\it finite} $r$, but in the northern hemisphere $r=\infty$ lies within
the AdS spacetime (see figure \ref{fig:accbh}). To transform to global AdS coordinates 
$\{R,\Theta\}$, one takes \cite{Podolsky:2002nk}
\be
1 + \frac{R^2}{\ell^2} = \frac{1 + (1-A^2\ell^2)r^2/\ell^2}{(1-A^2\ell^2)\Omega^2}
\qquad;\qquad
R \sin\Theta = \frac{r\sin\theta}{\Omega}
\ee 
the boundary, $R\to\infty$ now clearly corresponds to $\Omega\to0$, 
and the origin of Rindler coordinates corresponds to
$R_0 = A \ell^2/\sqrt{1-A^2\ell^2}$, i.e.\ the Rindler coordinates represent
those of an observer displaced from the origin of AdS, clearly requiring $A\ell<1$.
If $A\ell=1$, \eqref{rindler} simply becomes the Poincar\'e patch of AdS,
with a horizon appearing at $z = r\cos\theta = -\ell$. The C-metric
with $A\ell=1$ was considered in \cite{Emparan:1999wa}.

Adding in a black hole now introduces a conical singularity as described above,
as well as a black hole horizon $f(r_+)=0$, given by a rather unhelpful algebraic 
expression. If $A\ell>1$, the black hole will have an acceleration horizon at large 
$r$, and if $A\ell < 3\sqrt{3}/4\sqrt{2} \simeq 0.919$, there is only the black hole
event horizon present in the spacetime. 
However, as a consequence of picking the Hong-Teo coordinates and 
parametrisation, the boundary between having or not having
an acceleration horizon no longer lies at a single value of $A\ell$\footnote{Note
that in the traditional C-metric, the delineation between having an acceleration
horizon and not is $\tilde{A}\ell=1$ (considered in \cite{Emparan:1999wa}), however, 
this is not the same acceleration parameter as in the Hong-Teo description.}
but is instead determined by a mass dependent relation. 
For $3\sqrt{3}/4\sqrt{2}<A\ell<1$, there can be an
acceleration horizon beyond $r=\infty$ for suitable values of $mA<1/2$. 
Typically, in holographic applications, the conical deficit of the C-metric
is desired to be cloaked by such an horizon \cite{Hubeny:2009kz},
however, in our case we are interested in thermodynamics, so to
keep our thermodynamic variables unambiguous, we will take 
$A$ sufficiently small so that we have a single horizon, that of the black hole,
and thus a well-defined temperature corresponding to the periodicity
of Euclidean time at the regular Euclidean black hole horizon. We 
refer to this as a {\it slowly accelerating black hole}, and for clarity in plots
we will simply indicate this boundary as the value $A\ell=1$. 
A comprehensive exploration of the structure of AdS C-metric spacetimes, 
including their distortion of the boundary, was given in \cite{Krtous:2005ej}.

\subsection{Thermodynamics of the Slowly Accelerating C-metric}

In this section we perform a similar analysis to that of the string threading the 
black hole, but now we must allow for the tension of each of the conical deficits
of the C-metric to vary.
We start by finding the temperature and entropy of the black hole using
the conventional relations\footnote{See \cite{Herdeiro:2009vd}
for a discussion of the area law in the presence of conical singularities.}
\be
\beal
T &= \frac{f'(r_+)}{4\pi} = \frac{1}{2\pi r_+^2} \left [ (1-A^2 r_+^2) \left (
m - \frac{e^2}{r_+}\right) + \frac{r_+^3}{\ell^2(1-A^2 r_+^2)} \right]\\
S &= \frac{\rm Area}{4} =\frac{\pi r_+^2}{K(1-A^2 r_+^2)}
\eeal
\ee
checking the Smarr relation, we compute
\be
2TS =\frac{m}{K} - \frac{e^2}{K r_+^2}
+ \frac{r_+^3}{K\ell^2(1-A^2 r_+^2)^2} 
\ee
The charge of the black hole is given by \eqref{Qdef}, $Q = e/K$, and defining 
\be
V  = \frac{4\pi r_+^3}{3K(1-A^2r_+^2)^2}
\ee
as the modified thermodynamic volume, we obtain
\be
\frac{m}{K} = 2 TS  + Q\Phi_H - 2 PV\,.
\ee
Although it is tempting to identify $M=m/K$, this would be to ignore the asymptotics
of the spacetime. The experience of the rotating AdS black hole \cite{Gibbons:2004ai} is
that thermodynamic potentials should be normalised at infinity, and in the case
of rotation, the Boyer-Lindquist coordinates give a boundary that is rotating. Subtracting
off this rotation leads to an extra renormalisation of the thermodynamic mass, 
a correct Smarr formula and correct first law. 

Here, however, we cannot simply perform a similar electromagnetic
gauge transformation. Our electrostatic potential no longer
vanishes at infinity, and our boundary has an electric flux from pole to pole
\be
F = eA \sin\theta \, dt \wedge d\theta
\ee
We obviously cannot subtract this charge, as that would be a physical change, but
it does lead us to suspect that there may be a renormalization of electrostatic potential
and thermodynamic mass. We will show how to do this presently, but first consider
just the uncharged black hole, and consider variations in the position of the horizon
as in \eqref{simplevary}:
\be
\delta f(r_+) = f_+' \delta r_+  - 2 \frac{\delta m}{r_+} (1-A^2r_+^2)
- 2 A \delta A r_+ (r_+ - 2m) - 2 \frac{r_+^2}{\ell^3} \delta \ell = 0
\ee
The procedure is similar to the previous section, but we now have
more algebra involved in the variation of the thermodynamic parameters.
For example, in relating $\delta r_+$ to $\delta S$, we have:
\be
\delta S = \frac{2\pi r_+ \delta r_+}{K(1-A^2 r_+^2)^2} + 
\frac{2 \pi r_+^4 A\delta A}{K (1-A^2 r_+^2)^2} - 
\frac{\pi r_+^2}{(1-A^2 r_+^2)} \frac{\delta K}{K^2}
\ee
where our expressions for the tensions \eqref{tensions} give
\be
\frac{\delta K}{K^2} = 2 \left ( \delta \mu_+ + \delta \mu_- \right )
\quad ; \qquad \frac{m}{K} \delta A = - \left [
\delta \mu_+ - \delta \mu_- + A \delta (\frac{m}{K}) \right]\,.
\ee
Defining $M=m/K$, after some algebra one gets
\be
\delta M = V\delta P + T \delta S -
\delta \mu_+ \left [ \frac{r_+}{1+Ar_+} - KM \right ] -
\delta \mu_- \left [ \frac{r_+}{1-Ar_+} - KM \right ] \,.
\label{firstaccm}
\ee
Thus, the accelerating black hole has the same thermodynamic first law
as the non-accelerating black hole, but now with a thermodynamic length
for the piece of string attaching at each pole:
\be
\lambda_\pm = \frac{r_+}{1 \pm Ar_+} - KM
\ee
This obviously agrees with \eqref{TDlengthnoacc} for the string
threading the black hole, where $r_+$ has now been replaced by
$r_+/\Omega(r_+,\theta_\pm)$ at each pole. Note, a remarkably
similar expression was obtained by Kastor and Traschen 
\cite{Kastor:2012dt} in the context of a ``Kaluza-Klein-like'' gravitational
tension (see e.g.\ \cite{Traschen:2001pb,Harmark:2004ch})
associated to the rotational symmetry of the black hole.

Now let us consider the addition of charge. Following the same procedure of
varying the horizon as before leads to the relation
\be
\delta (\frac{m}{K}) = T \delta S + V \delta P + \Phi_H\delta Q
- \frac{r_+\delta \mu_+}{1+Ar_+}- \frac{r_+\delta \mu_-}{1-Ar_+}
+ \frac{m\delta K}{2K^2}
\ee
where now our expressions for the tensions lead to
\be
\beal
\frac{m}{K} \delta A &= -\delta \mu_+ + \delta \mu_- 
- A \delta \left (\frac{m}{K} \right) \\
[1 - e^2 A^2 ] \frac{\delta K}{2K^2} &= A^2 e \delta Q
- \frac{e^2 A^2}{m} \delta \left ( \frac{m}{K} \right )
+ \delta \mu_+ \left [ 1 - \frac{Ae^2}{m} \right ] + \delta \mu_- \left [ 1 + \frac{Ae^2}{m} \right ]
\eeal
\label{deltaKcharge}
\ee

Keeping an open mind, we define our thermodynamic mass and electrostatic
potential as:
\be
M = \gamma(e,A) \frac{m}{K} \quad ; \qquad
\Phi = \Phi_H - \Phi_0
\ee
where $\Phi_0$ is a correction, re-zeroing the potential, analogous to the correction of
the angular potential of the Kerr-AdS black hole, but without the corresponding 
interpretation of being the value of the original potential at infinity.

Next, we compare
\be
\beal
\delta M &= \gamma \delta \left ( \frac{m}{K} \right ) 
+ \frac{m}{K} (\gamma_e \delta e + \gamma_A \delta A)
\\&= \gamma \delta \left ( \frac{m}{K} \right ) 
+ m \gamma_e \delta Q + m e \gamma_e \frac{\delta K}{K^2}
+ \gamma_A \frac{m}{K} \delta A
\eeal
\ee
to
\be
\beal
T \delta S + V \delta P + \Phi \delta Q -\lambda_+ \delta \mu_+ - \lambda_-\delta \mu_-
= \delta \left ( \frac{m}{K} \right ) 
- \Phi_0 \delta Q - \frac{m\delta K}{2K^2}&\\
+ \left [ \frac{r_+}{1+Ar_+} - \lambda _+ \right]\delta \mu_+
+ \left [ \frac{r_+}{1-Ar_+} - \lambda _- \right]\delta \mu_-&
\eeal
\ee
where $\lambda_\pm$ are to be determined. After some
algebra, we obtain
\be
\beal
\delta M &- T \delta S - V \delta P - \Phi \delta Q 
+ \lambda_+ \delta \mu_+ + \lambda_-\delta \mu_-\\
&=
\left [ (1-e^2 A^2) \gamma-2 e^3A^2 \gamma_e - A(1-e^2 A^2) \gamma_A -1 \right] 
 \frac{ \delta \left ( {m/K} \right ) }{(1-e^2A^2)}  \\
&+ \left [ m (1+e^2 A^2) \gamma_e + mA^2 e
+(1-e^2 A^2) \Phi_0 \right] \frac{ \delta Q }{(1-e^2A^2) }\\
&+ \left [ \lambda_+ - \frac{r_+}{1+Ar_+} - \gamma_A 
+ \frac{(2e\gamma_e+1)}{1-e^2 A^2} \left ( m - e^2A \right) 
\right] \delta \mu_+\\
&+ \left [ \lambda_- - \frac{r_+}{1-Ar_+} + \gamma_A 
+\frac{(2e\gamma_e+1)}{1-e^2 A^2} \left ( m + e^2A \right) 
\right] \delta \mu_-
\eeal
\ee
for our first law to hold, clearly the RHS of this equation must vanish, leading to 
a constraint for $\gamma$:
\be
(1-e^2 A^2) \gamma-2 e^3A^2 \gamma_e - A(1-e^2 A^2) \gamma_A =1
\quad \Rightarrow \qquad \gamma = \frac{1}{1+e^2 A^2}
\ee
thus specifying our thermodynamic mass
and determining $\Phi_0$ and $\lambda_\pm$:
\be
\beal
M &= \frac{m}{K(1+e^2 A^2)} \\
\Phi_0 &= \frac{meA^2}{1+e^2 A^2}\\
\lambda_\pm &= \frac{r_+}{1\pm Ar_+} - \frac{m(1-e^2A^2)}{(1+e^2 A^2)^2} 
\mp \frac{e^2A}{(1+e^2 A^2)}
\eeal
\label{TDparams}
\ee
This is a rather unusual set of relations, the off-set of the electrostatic potential
depends on mass, and the thermodynamic mass depends on charge. We view
this as a consequence of the fact that for the accelerating black hole, the electric
potential cannot be gauged away at infinity -- there is a polar electric field at the
AdS boundary, thus mass and charge are inextricably intertwined.

\section{The Thermodynamic Length}

To recap, we have shown that allowing for a varying conical deficit
in black hole spacetimes, the first law of thermodynamics becomes
\be
\delta M = T \delta S + V\delta P + \Phi \delta Q - \lambda_+ \delta \mu_+
- \lambda_- \delta \mu_-
\ee
where the relevant thermodynamical variables are given in \eqref{TDparams}.
In order to accommodate varying 
tension, we have to define a {\it thermodynamic length},
\be
\lambda_\pm = \frac{r_+}{1\pm Ar_+} - \frac{m(1-e^2A^2)}{(1+e^2 A^2)^2} 
\mp \frac{e^2A}{(1+e^2 A^2)}
\ee
for each conical deficit emerging from each pole. This length consists of 
a direct geometrical part, a mass dependent correction, and finally, a shift
in the presence of charge.

It is interesting to compare this mass-dependent shift of the thermodynamic
length to the correction of the thermodynamic volume for a rotating black
hole \cite{Cvetic:2010jb,Dolan:2011jm}:
\be
V = \frac{4\pi}{3} \left \{ \frac{r_+(r_+^2+a^2)}{K} + a^2 M \right\}
\ee
In this case, the first term is the expected geometric volume of the 
interior of the black hole, the second term being a rotation dependent
correction. With this appropriately shifted thermodynamic volume,
the black hole always satisfies a {\it Reverse Isoperimetric Inequality}
\cite{Cvetic:2010jb}, that is to say, the entropy of a black hole of a given
thermodynamic volume is always maximised for the purely spherical
(Schwarzschild) black hole. (We verified in \cite{Appels:2016uha} that
the accelerating black hole satisfies the reverse isoperimetric inequality;
allowing tension to vary does not alter this result.)

\begin{figure}
\center{\includegraphics[scale=0.7]{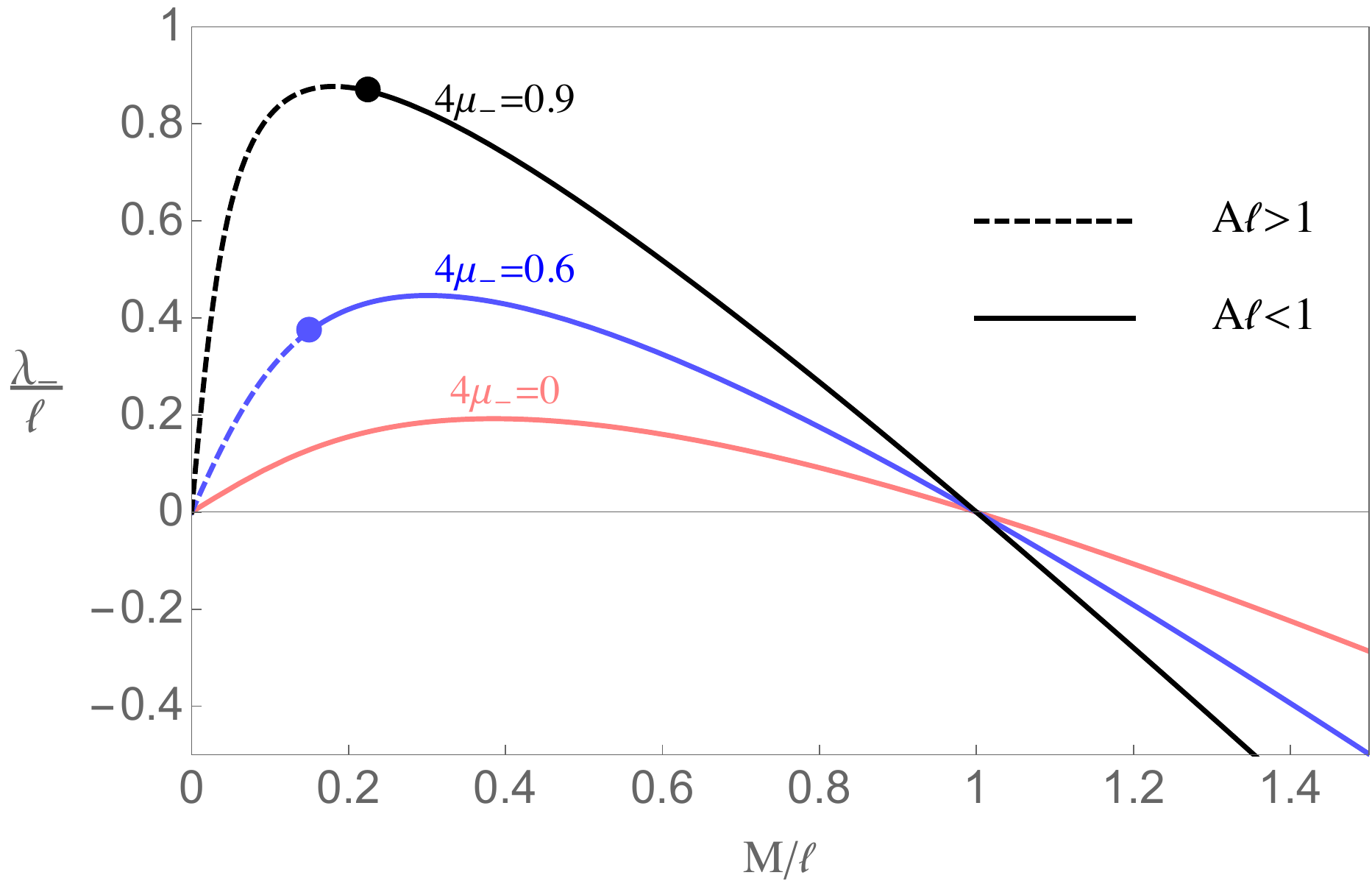}}
\caption{A plot of thermodynamic length of the south pole tension for regular
north pole, $\mu_+=0$. The limit of the Schwarzschild-AdS black hole is shown
in pink ($\mu_-=0$). The slowly accelerating r\'egime, $A\ell<1$, is shown as a 
solid line, and $A\ell>1$ is shown dashed. 
}
\label{fig:TDlength}
\end{figure}

Notice that the correction term for this thermodynamic volume is
always positive, whereas the correction term for thermodynamic 
length is actually negative.
This means that for large enough mass, the thermodynamic length
itself becomes negative, as shown in figure \ref{fig:TDlength} for 
an uncharged black hole. The picture for a charged black hole is 
similar, although the critical value of $M$ for which $\lambda_-$ 
becomes negative is larger.

From figure \ref{fig:TDlength}, we see that the thermodynamic length 
becomes negative for `large'
black holes, i.e.\ those for which the thermodynamic mass is of similar
order (or higher) than the AdS scale. Setting this in the context of the
`cosmic string' capture process considered in section \ref{sec:singlemu} 
for the vacuum black hole, this would mean that the thermodynamic mass 
must increase during a capture, as entropy cannot decrease. This seems
at first counter to the notion that the string itself does not carry `ADM' 
mass, however, the heuristic argument of section \ref{sec:singlemu}
relies somewhat on the notion that a cosmic string and black hole can
be sufficiently separated so that one can consider their thermodynamical
(and other) properties independently. For large black holes in AdS this
is manifestly not the case.

It is also interesting to compare these results for varying tension to our previous 
work \cite{Appels:2016uha}, where $K$ and $\mu_\pm$ were held fixed. 
With these assumptions, $eA$ and $mA$ were fixed, however, we did not alter the 
thermodynamic mass from $m/K$, nor the electrostatic potential 
from $\Phi_H$. The two sets of results are consistent, since
\be
\Phi_0 Q = \frac{m e^2 A^2}{K(1+e^2 A^2)} 
= \frac{m}{K} \left [ 1 - \frac{1}{1+e^2 A^2} \right] = \frac{m}{K} - M
\ee
The shift in electrostatic potential multiplied by charge therefore
balances the shift in thermodynamic mass in both the Smarr formula,
and indeed the first law with the assumptions made in \cite{Appels:2016uha} 
since $eA$ was required to be fixed. However, it is worth revisiting
these assumptions in the light of our work here on varying tension.

First, notice that our charged C-metric has parameters: $m$, relating to the mass
of the black hole, $e$ to its charge, $A$ to its acceleration, and $K$, that relates to
an overall conical deficit. $K$ is the one parameter that has no immediately obvious
physical interpretation, indeed seems more like a coordinate choice, thus fixing
$K$ was natural. However, now armed with our better understanding of the metric
and its thermodynamics, we see that in fixing the tensions of the deficits, we are fixing
two physical quantities, thus we should only find that {\it two} combinations of
the solution parameters are fixed. Therefore, we should not fix $K$ a priori, but instead
just the combinations of parameters that fix the tensions:
\be
\beal
2(\mu_++\mu_-) &= 1 - \frac{1+e^2A^2}{K} \\
\mu_+ - \mu_- &= - \frac{mA}{K}
\eeal
\ee
From these expressions, we see that if charge vanishes, then indeed fixing
tensions fixes $K$ and the combination $mA$, but if charge does not vanish,
then we can no longer conclude that $\delta K=0$. Instead
\be
\frac{\delta K}{K} = \frac{\delta (mA)}{mA} = 2\frac{eA\delta (eA)}{1+e^2A^2}
\ee
i.e.\ we have {\it two} constraints on the variation of our parameters. 
Thus, for example if we throw a small mass $m_0$ into the black hole, we expect
$\delta M = m_0$, $\delta Q=\delta P = 0$. Using the expression for $M$ and
the tensions we then find:
\be
\beal
\frac{\delta K}{K^2} &= - 2 \frac{e^2 A^2}{m}\delta M &
\delta A &= - (1-e^2 A^2) \frac{AK}{m} \delta M\\
\delta m &= (1- 3 e^2 A^2) K \delta M&
\delta e &= -2\frac{e^3A^2K}{m} \delta M
\eeal
\ee
indicating that the acceleration of the black hole drops, as expected.

\section{Critical Behaviour of Accelerating Black Holes}

Given that we are working in anti-de Sitter spacetime, we can ask whether
there is something analogous to a Hawking-Page phase transition 
\cite{Hawking:1982dh} for our accelerating black holes, although it is difficult to see
how one could actually have a phase transition between a system with a conical
deficit along one polar axis only, and a presumably totally regular radiation bath. 
However, recall that a black hole
in AdS behaves similarly to a black hole in a reflecting box, with the negative
curvature of the AdS providing the qualitative reflection. For small black holes,
the effect of the negative curvature is sub-dominant to the local curvature
of the black hole, and the black hole has negative specific heat, as in the
vacuum Schwarzschild case. For black holes larger than the AdS radius, the
vacuum curvature dominates, and the black hole has positive specific heat,
in particular, there is a minimum temperature for a black hole in AdS, below this
temperature, only a radiation bath can be a solution to the Einstein equations at
finite $T$. Plotting the Gibbs free energy as a function of temperature shows 
both the allowed states, as well as the preferred one for a given temperature.
At very low $T$, the only allowed state is a radiation bath. Above a critical 
temperature $T_c = \sqrt{3}/2\pi\ell$, one can have either a radiation bath,
or a black hole (that may be either `small' or `large'). However for $T>1/\pi\ell$, 
the large black hole is not only thermodynamically stable (in the sense
of positive specific heat) but thermodynamically preferred, and a radiation bath
will spontaneously transition into a large black hole. 

\begin{figure}
\center{\includegraphics[scale=0.58]{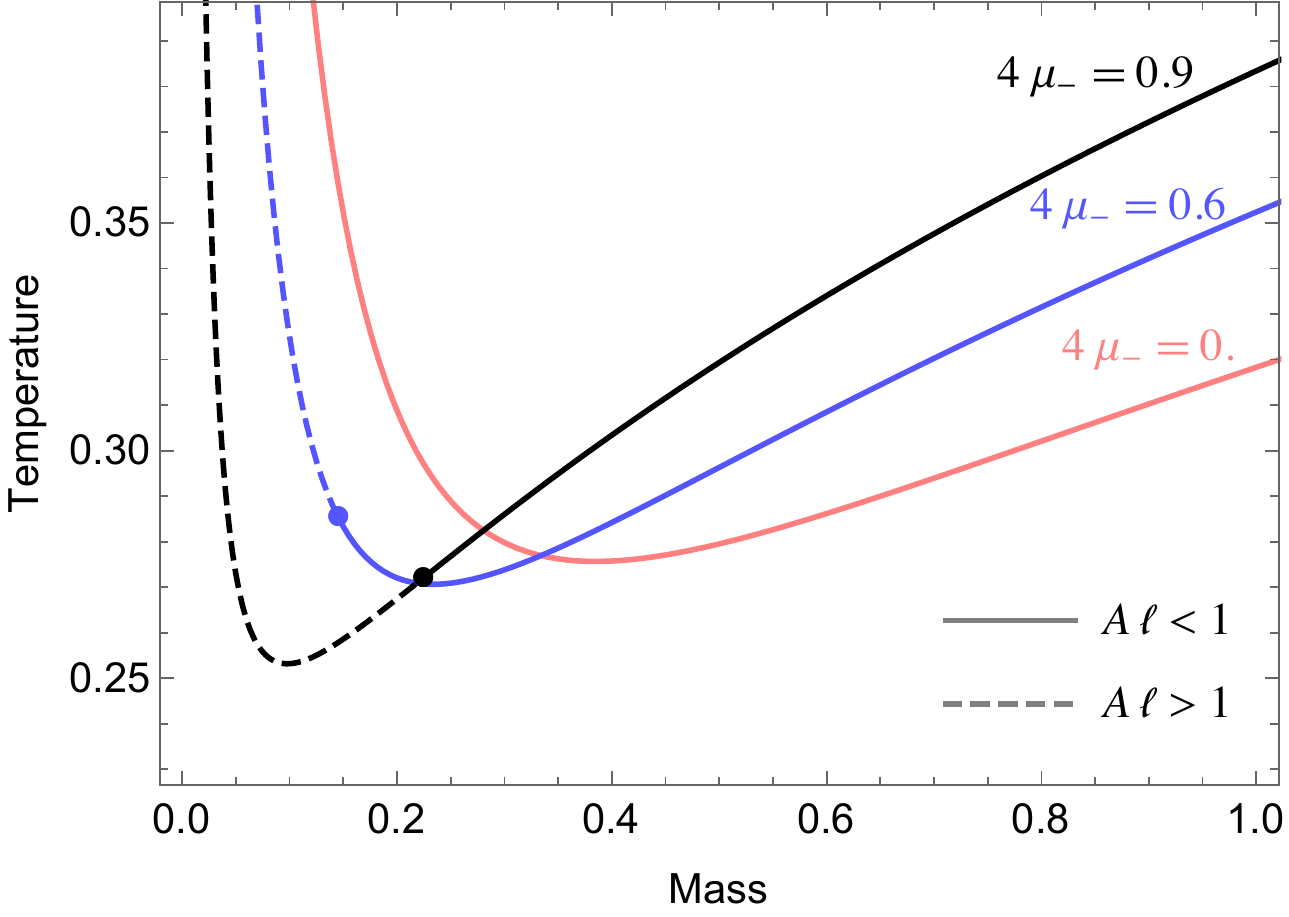}~
\includegraphics[scale=0.6]{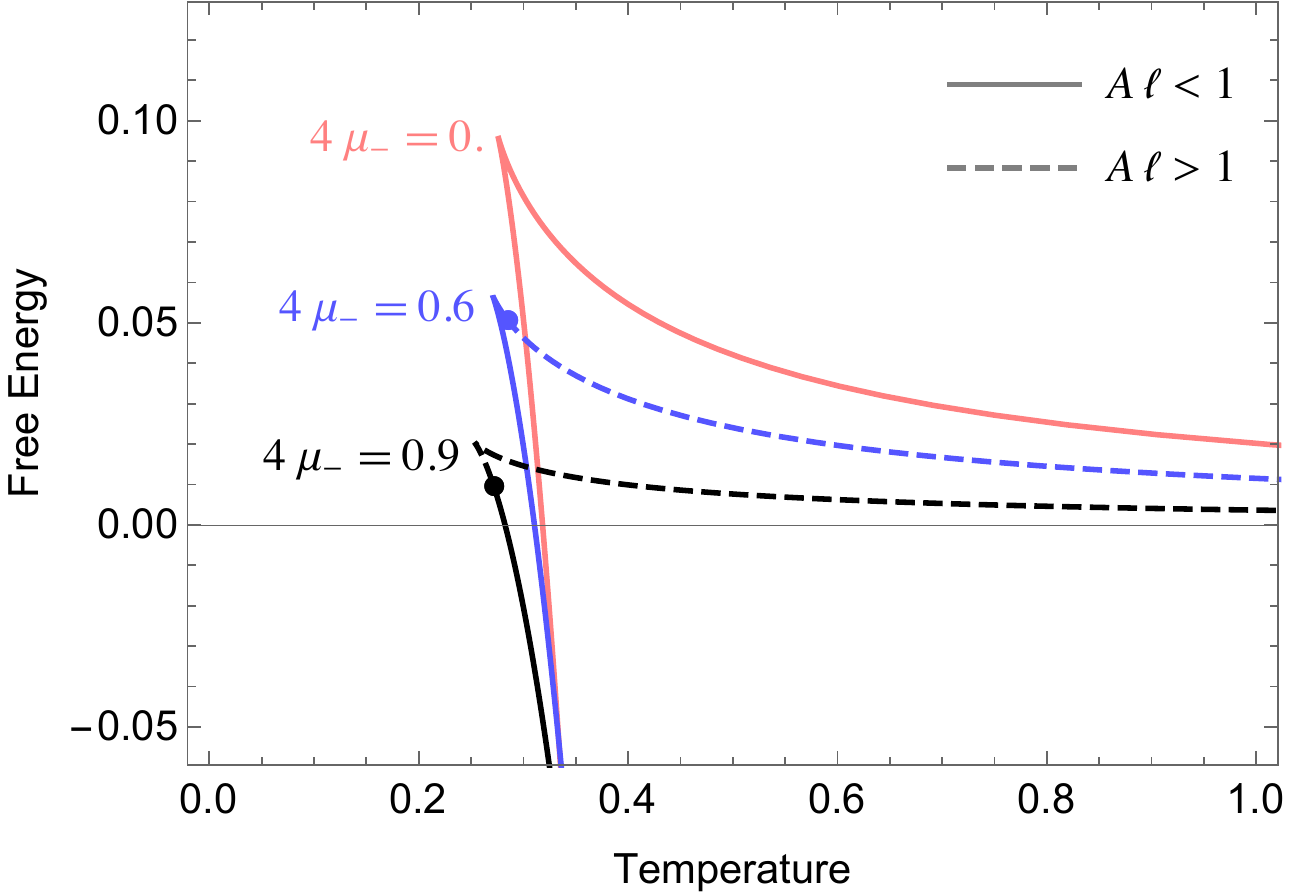}}
\caption{{\bf Left:} A plot of temperature as a function of mass 
(in units of $\ell$) for the uncharged black hole. 
The slowly accelerating r\'egime is shown as a solid line, and the inferred 
local horizon temperature for $A\ell>1$ is shown dashed. Note how for larger
string tension (hence greater acceleration) the region of positive specific heat 
increases. {\bf Right:} A similar plot, but now showing the Gibbs free energy as a function
of temperature.
}
\label{fig:TvMnocharge}
\end{figure}

First consider the situation where our accelerating black hole is uncharged.\footnote{
Note: in all explicit examples and figures in this section we take the $\theta=0$
axis to be regular ($\mu_+=0$). This is for simplicity, including a nonzero
north pole tension does not alter the essential physics of what we present here.}
Fixing the tension of the string, we can plot the temperature of our black hole
as a function of its mass, $M$, as shown in figure \ref{fig:TvMnocharge}.
This figure shows how increasing acceleration actually makes a black hole
of given mass {\it more} thermodynamically stable in the sense of positive specific
heat. Figure \ref{fig:TvMnocharge} also shows the corresponding Gibbs free energy,
indicating the would-be Hawking-Page transition occurs at lower temperatures as
acceleration increases.

At first sight, this is rather curious, as a naive examination of the uncharged 
C-metric shows that the Newtonian potential, $f(r)$ has the cosmological 
constant ameliorated by the
acceleration: $f(r) = r^2 (1 / \ell^2 - A^2) \simeq r^2/\ell_{\rm eff}^2$. Given
that one often imagines that it is the black hole radius relative to the 
confining `box' of AdS that is causing the thermodynamic stability of the
large black holes, this looks rather confusing: increasing acceleration appears
to counteract the AdS length scale. However, this intuition is too naive: the
relevant effect is the spacetime curvature in the vicinity of the horizon, and
whether the black hole or the cosmological constant is dominant (larger
black holes having smaller tidal forces). Computing the Kretschmann scalar
at the event horizon demonstrates that indeed, increasing acceleration for
a given mass lowers the local tidal forces due to the black hole. 
In fact, it is easy to compute the ``Hawking-Page'' transition temperature,
assuming the radiation bath to have zero Gibbs energy from the expressions
for $TS$ and $M$ in terms of $r_+$, $A$ and $\ell$. A brief calculation gives
\be
\beal
T_{HP} (r_+, \ell, A) &= \frac{1}{4\pi r_+ }\left [
\frac{ 3r_+^2}{\ell^2 ( 1- A^2 r_+^2) } + 1 \right ]\\
& \simeq \frac{1}{2\pi \ell} \left ( 1 - \frac32 A^2 \ell^2 + {\cal{O}} (A^4 \ell^4) \right)
\;.
\eeal
\ee
While the acceleration parameter $A$ is not a thermodynamic charge, instead
being related to the tension via $M$, nonetheless, the general picture is that
increasing tension increases acceleration, thereby decreasing the temperature at which
the ``Hawking-Page'' transition occurs.

Now consider adding a charge to the black hole, for which we might now expect
a richer phase structure, possibly with critical phenomena analogous to the 
isolated charged AdS black hole 
\cite{Chamblin:1999tk,Cvetic:1999ne,Chamblin:1999hg}. The critical phenomena
occur due to the three possible phases of black hole behaviour for varying mass.
In the presence of charge, there is now a lower limit on the mass parameter of 
the black hole, set by the extremal limit where the temperature vanishes. 
Increasing the mass of the black hole moves it away from extremality, thus
increasing temperature, rendering the specific heat positive near 
this lower limit. For large mass black holes, we are also in a positive specific heat
r\'egime where the local vacuum curvature is dominant in the near horizon geometry.
Depending on the size of the charge relative to the vacuum energy, there can be
an additional negative specific heat r\'egime where the black hole is small enough 
that its local curvature is dominant, but is far enough from extremality that the usual
Schwarzschild negative specific heat type of behaviour pervades. 
Given that for uncharged accelerating
black holes, increasing tension lowers the critical temperature at which the 
transition to positive specific heat occurs, we expect this `swallowtail' behaviour
to be mitigated for charged accelerating black holes in the canonical ensemble,
and indeed this is what is observed.
\begin{figure}
\center{\includegraphics[scale=0.7]{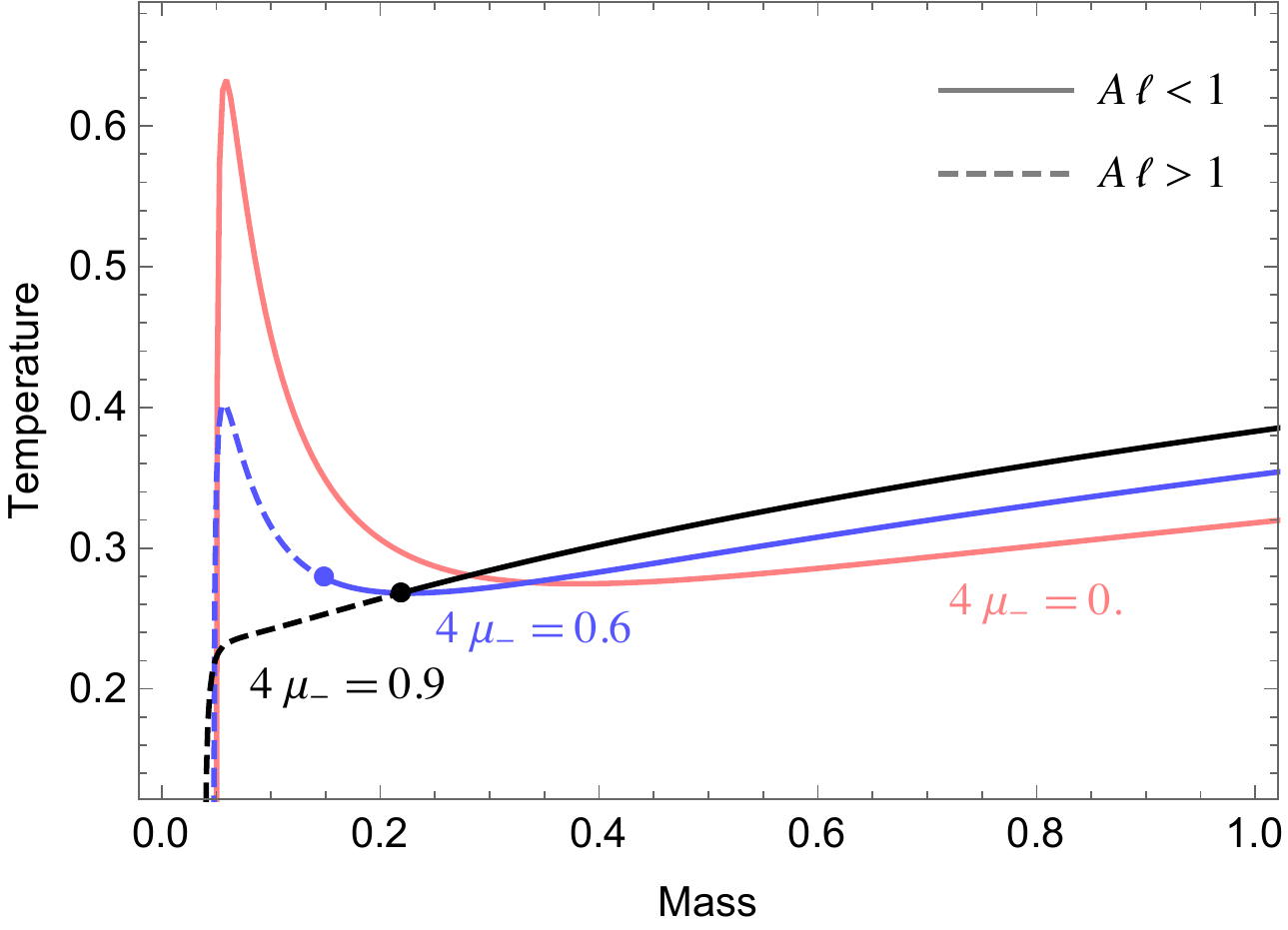}}
\caption{A plot of temperature as a function of mass for the charged black hole,
with fixed $Q=0.05\ell$, and varying tension as labelled. As before, the
slowly accelerating r\'egime is shown as a solid line, and $A\ell>1$ is shown dashed. 
}
\label{fig:TvMcharge}
\end{figure}

\begin{figure}
\center{\includegraphics[scale=0.59]{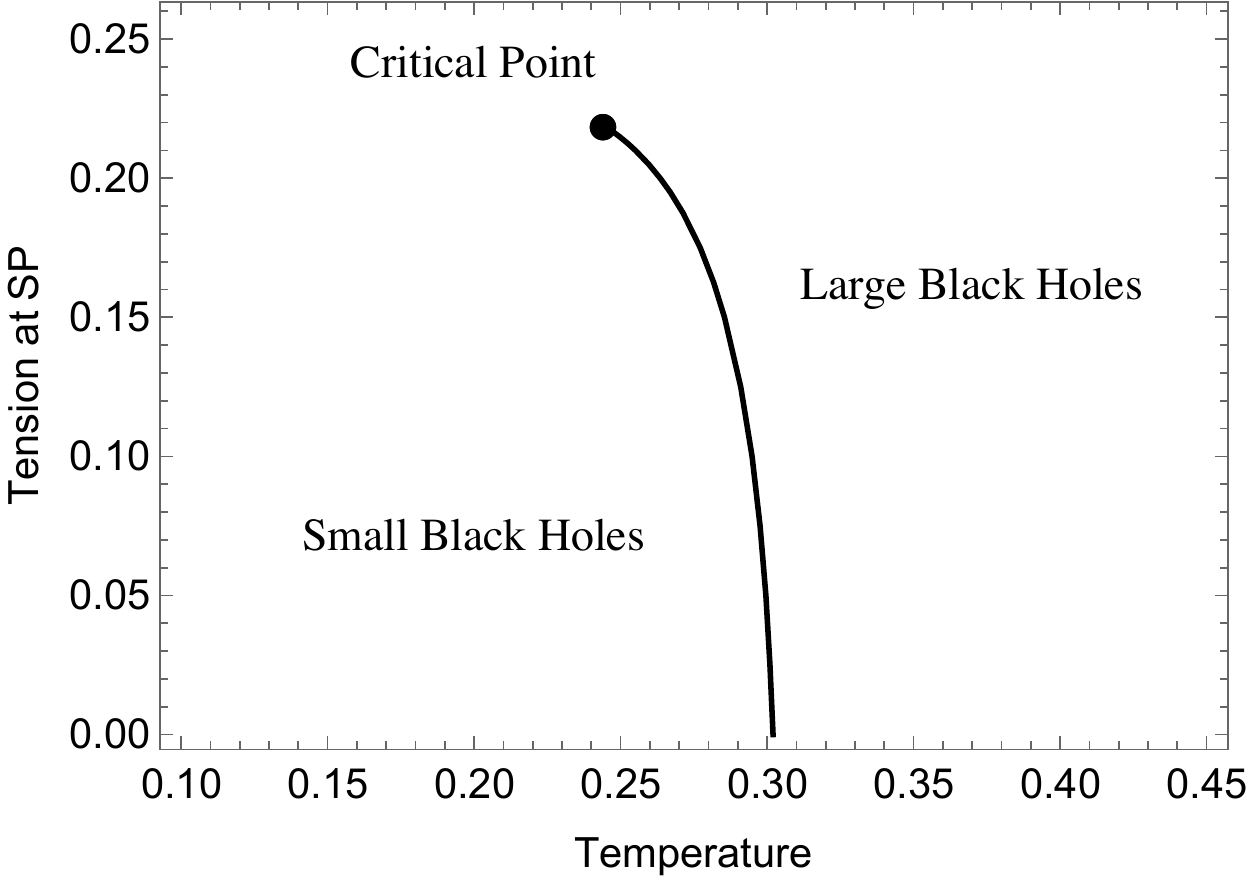}~
\includegraphics[scale=0.57]{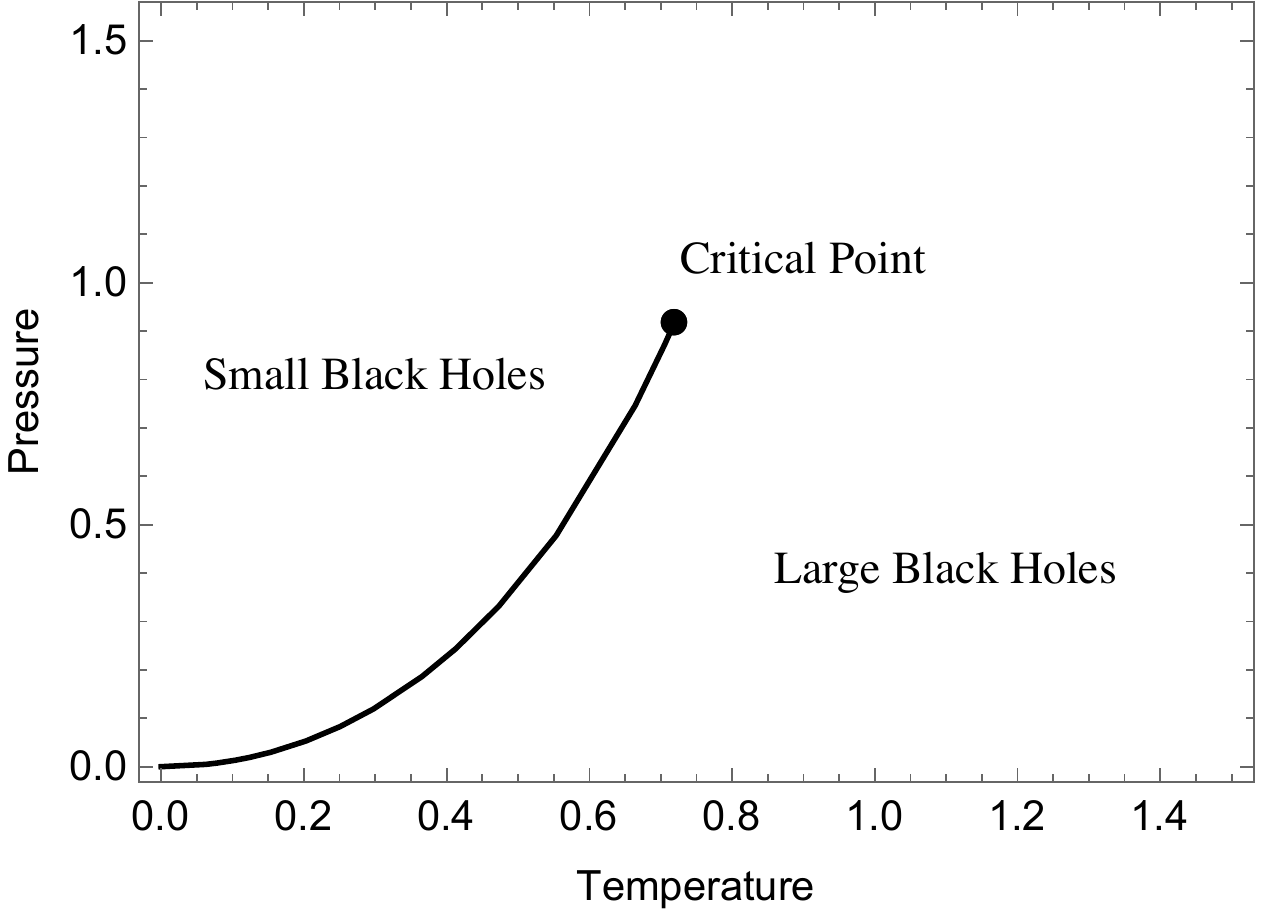}}
\caption{The coexistence line for the charged black hole shown for 
varying tension and cosmological constant with the black hole charge
is fixed at $Q=0.05$. On the left, $\ell=1$, and the value of tension at the
critical point is $\mu_c = 0.219$. On the right, $4\mu_- = 0.3$, and the
critical value of the AdS radius is $\ell_c = 0.36$.
}
\label{fig:coexist}
\end{figure}

We first explore the accelerating black hole in the canonical ensemble, i.e.\
where the charge, $Q$, of the black hole is fixed, but we allow $M$ and $\mu_-$
to vary.
In figure \ref{fig:TvMcharge}, we give a representative plot of temperature 
as a function of black hole mass for $Q=0.05\ell$ to illustrate how increasing 
tension gradually removes the negative specific heat phase of the black hole.
\begin{figure}
\center{\includegraphics[scale=0.58]{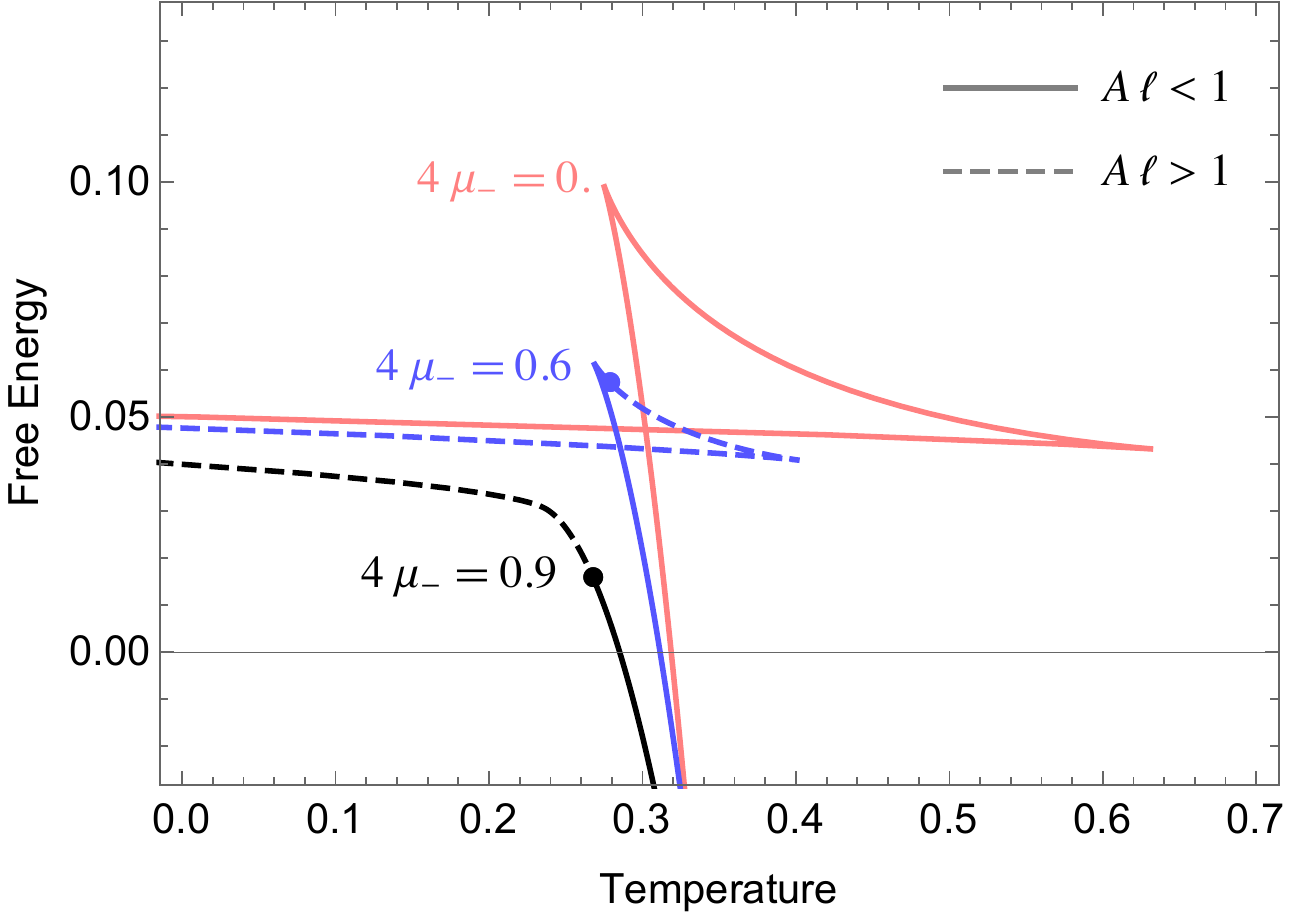}~
\includegraphics[scale=0.58]{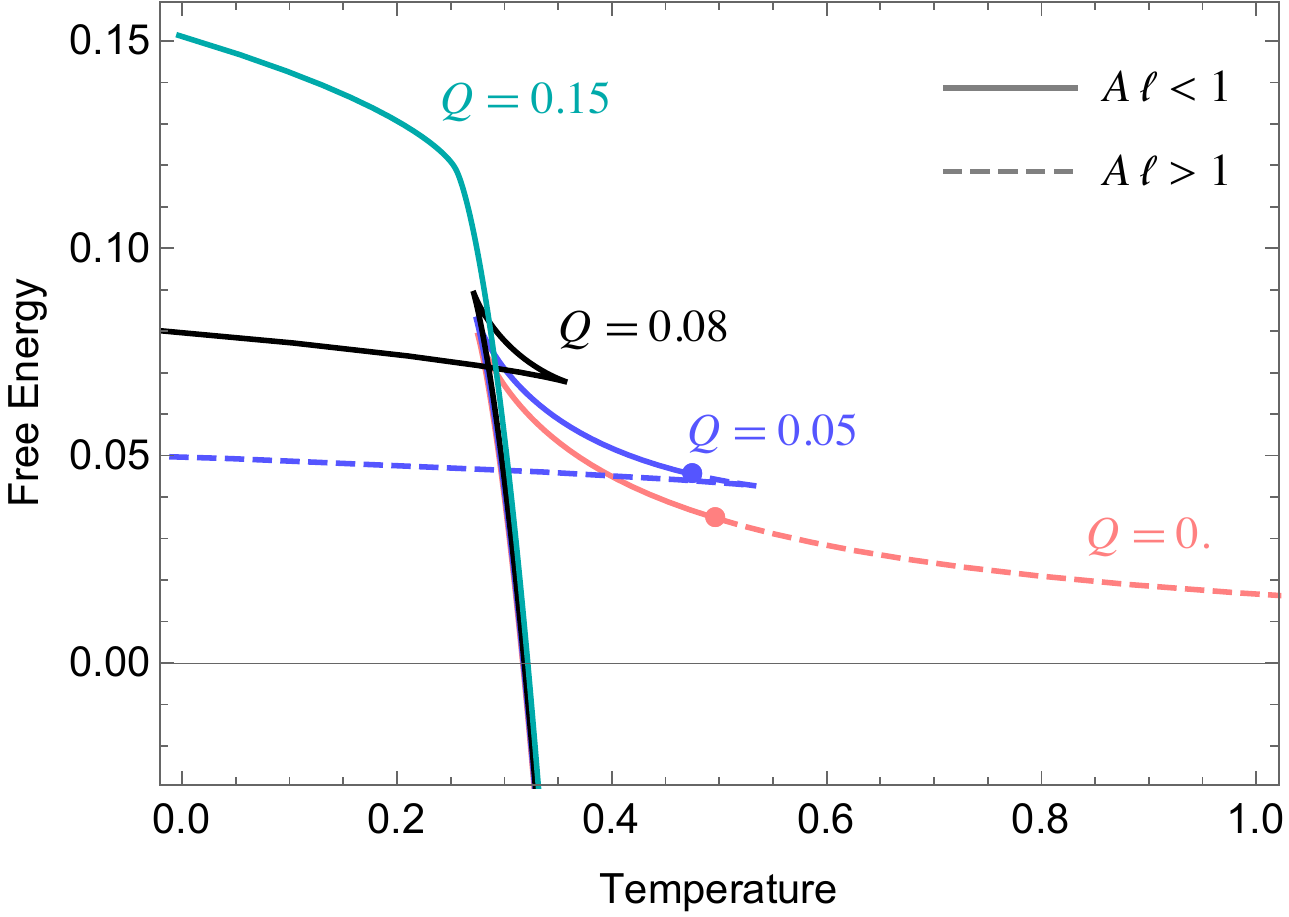}}
\caption{A plot of the free energy as a function of temperature for 
varying tension with $Q=0.05\ell$ on the left, and varying charge with $4\mu_-=0.3$
on the right. 
}
\label{fig:GvTcharge}
\end{figure}
Figure \ref{fig:GvTcharge} shows the variation of the free energy 
$F=M-TS$ with temperature for varying tension and charge. As tension 
is increased, the swallowtail becomes smaller, and eventually disappears, 
analogous to the situation where the charge is gradually increased, shown 
on the right in figure \ref{fig:GvTcharge}. The free energy plot
tells us that at low temperatures, we have the near extremal black hole,
however as the mass of the black hole increases there is a critical value
at which there is a spontaneous transition to a larger black hole with positive
specific heat. The existence of this transition relies on the presence of
the intermediate region of negative specific heat for the charged black hole.
For large enough tension (or charge relative to $\ell$), there is a critical point
at which this intermediate
r\'egime disappears, and the phase transition along with it. 
Figure \ref{fig:coexist} shows the ``van der Waals'' like behaviour of 
this coexistence curve for varying tension (in analogy to the varying
potential plots of \cite{Chamblin:1999tk}),
and cosmological constant (in analogy to \cite{Kubiznak:2012wp}).

\begin{figure}
\center{\includegraphics[scale=0.57]{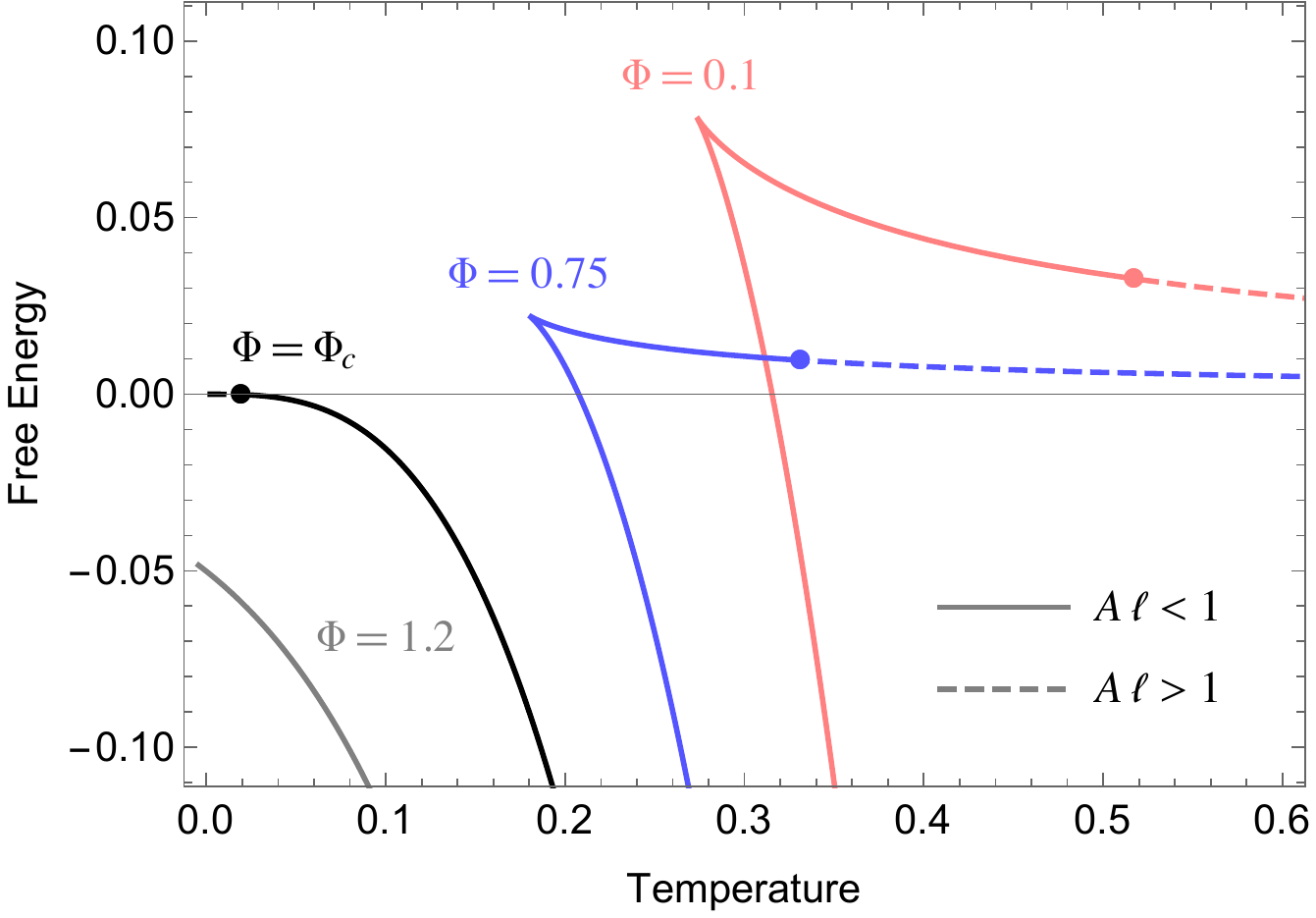}~
\includegraphics[scale=0.57]{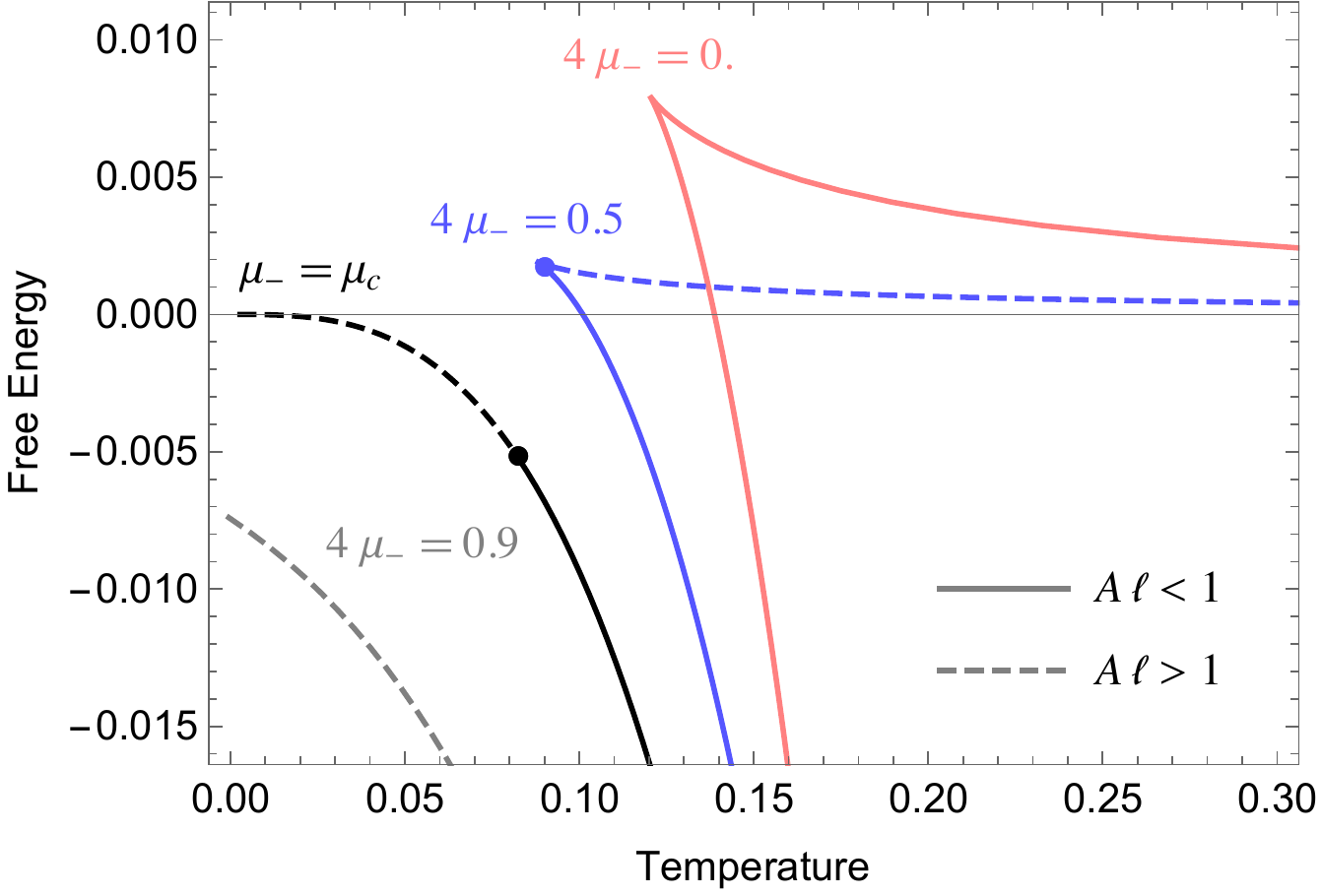}}
\caption{The Gibbs potential in the grand canonical ensemble
as a function of temperature, on the left with $4\mu_- = 0.3$
for varying potential as labelled, and on the right with $\Phi=0.9$
and varying tension as labelled in the plot.
}
\label{fig:grand}
\end{figure}

\begin{figure}
\center{\includegraphics[scale=0.75]{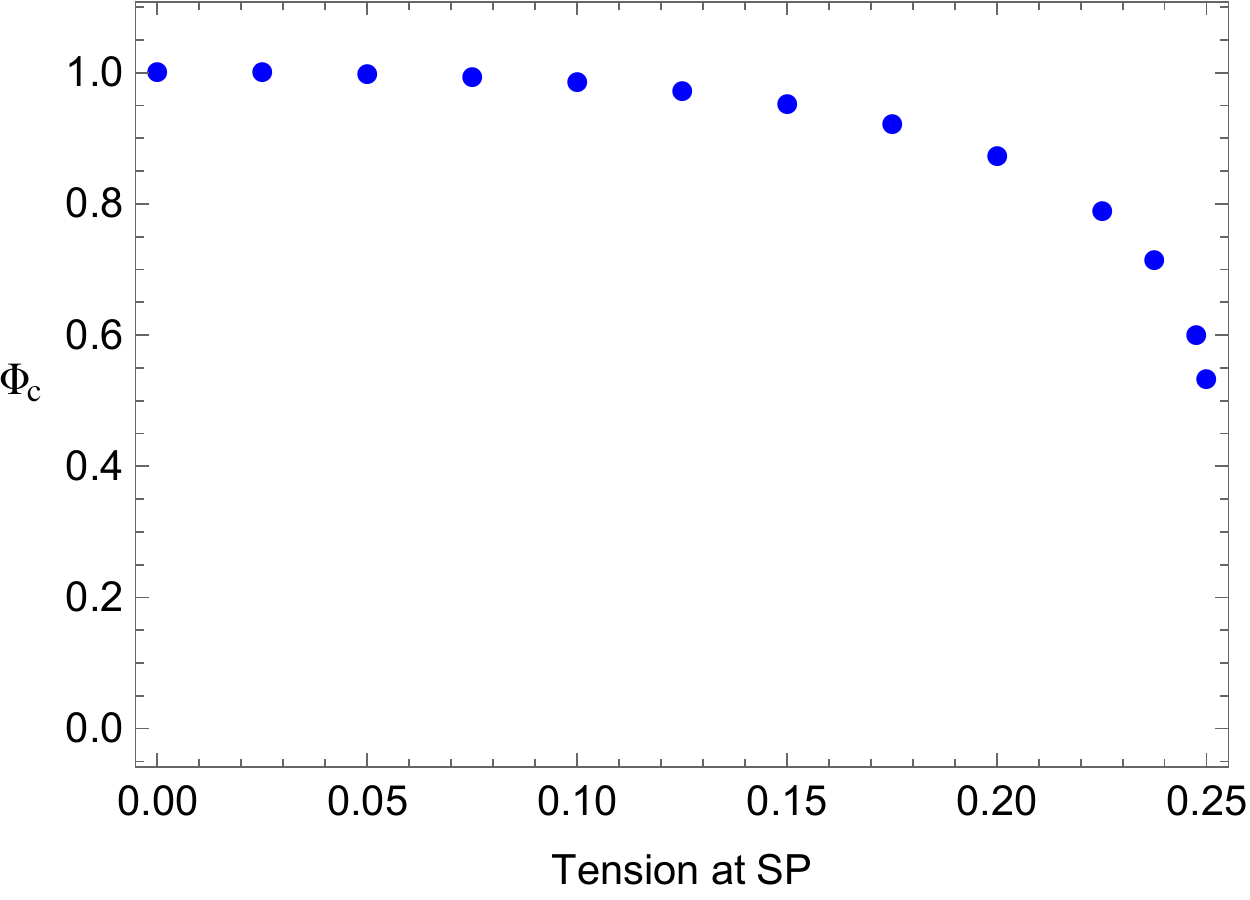}}
\caption{A plot of the critical value of $\Phi_c(\mu_-)$ at which a black hole
is always preferred for all temperatures as a function of the tension. 
}
\label{fig:phicrit}
\end{figure}

Finally, for completeness we consider the thermodynamics of the
accelerating charged black hole in the grand canonical ensemble, 
where we now allow charge to vary. The Gibbs potential is now
$G=M-TS-Q\Phi$, with 
\be
\Phi = \Phi_H - \Phi_0
= \frac{e}{r_+} - \frac{meA^2}{1+e^2A^2} 
\label{grandcanphi}
\ee 
kept fixed. The interesting feature of fixed potential, as noted 
in \cite{Chamblin:1999tk} for an isolated RNAdS 
black hole, is that there is a critical value of $\Phi$
delineating two qualitatively different behaviours of the black hole.
For small fixed potentials, the charged AdS black hole can never approach
extremality. This can be seen by noting that $f=f'=0$ at extremality, where
$f(r)$ is the RNAdS black hole potential. Solving these
algebraic equations, and substituting $\Phi_{RN} = e/r_+$, one finds
the constraint $3r_+^2 /\ell^2= \Phi_{RN}^2-1$, thus for $|\Phi_{RN}|<1$ 
there is no possibility of extremality. In our case, for the charged
accelerating black hole, the algebraic relations for extremality at 
fixed potential are considerably more complicated partly due to the 
extra acceleration parameter, but mostly because of the complicated
expression for $\Phi$ \eqref{grandcanphi}. However, the same
principle applies, and we also observe a similar phase transition
from small to large $\Phi$, where the critical value of $\Phi$ is now
tension dependent. Figure \ref{fig:grand} demonstrates this behaviour 
showing the analogous plot to \cite{Chamblin:1999tk} with acceleration
for fixed $\mu_-$, and also how the behaviour depends on $\mu_-$ at 
fixed $\Phi$, illustrating how increasing
$\mu_-$ improves the thermodynamic viability of the black hole.
Figure \ref{fig:phicrit} shows how the critical value of the potential,
where only positive specific heat black holes are allowed,
varies with tension.

\section{Conclusion}

To sum up: we have shown how to allow for a varying conical deficit
in black hole spacetimes, and found the relevant thermodynamical 
variables to describe the system. The appropriate first law has a 
varying tension term, with a thermodynamic length as its potential. 
This length consists of a direct geometrical part, a mass dependent correction, 
and finally, a shift in the presence of charge. The thermodynamic phases of
accelerating black holes exhibit similar behaviour to their non-accelerating 
AdS cousins, however, the impact of acceleration is to improve the thermodynamic
stability of the black holes.

It is interesting to note that the first law indicates that if the tension of a defect
is fixed, then there is no contribution to the variation of $M$ coming from 
tension, yet, if the black hole increases its mass and hence its horizon radius, 
the horizon will now have consumed a portion of the string along each pole.
This does not appear in the thermodynamic relation. This reinforces the
interpretation of $M$ as the {\it enthalpy} of the black hole \cite{Kastor:2009wy}.
Although the black hole increases its internal energy by swallowing some cosmic string,
it has also displaced the exact same amount of energy from the environment,
resulting in no net overall gain in the total energy of the thermodynamic system
(other than the mass that was added to the black hole in the first place).

It would be interesting to consider whether there are any holographic applications
for these solutions. (See \cite{Anber:2008zz} for a discussion of the CFT stress-energy
at the background Rindler temperature.)
Typically, one avoids having such distorted boundaries, however,
the fact that the thermodynamics of these systems is now well defined perhaps
suggests this is worth a second look. 

Finally, we have not discussed rotating accelerating black holes here. Even for
the non-accelerating rotating black hole, the thermodynamics are more subtle,
requiring an adjustment of the angular velocity \cite{Gibbons:2004ai}
in order to define it relative to a non-rotating frame at infinity. For a conical defect
running through the black hole (though see \cite{Gregory:2013xca,Gregory:2014uca}
for a full discussion of subtleties of replacing deficits by finite width defects)
one can follow the method of section \ref{sec:singlemu} to generalise the appropriate
thermodynamical variables of Gibbons et al.\ \cite{Gibbons:2004ai} to:
\be
\beal
M &= \frac{m}{KL} \;\;\;; \;\;\;\;&
V &= \frac{4\pi}{3} \left [ \frac{r_+(r_+^2+a^2)}{K} + a^2 M \right]\\
J &= \frac{am}{K^2} \;\;\;; \;\;\;\;&
\Omega &= \Omega_H-\Omega_\infty
= \frac{aK}{r_+^2+a^2} + \frac{aK}{\ell^2 L} \;\;\;; \;\;\;\;\\
Q &= \frac{e}{K} \;\;\;;\;\;\;\;& \Phi &= \frac{er_+}{r_+^2+a^2}
\eeal
\ee
where $K$ is the (arbitrary) parameter determining the periodicity of the azimuthal angle
as before, and $L$ is a new parameter due to black hole rotation:
\be
L = 1 - \frac{a^2}{\ell^2}
\ee
In \cite{Gibbons:2004ai}, $K=L$ was mandated by having no conical deficit on the
rotation axis, however, here we are being more general. Allowing the tension of any
deficit to vary leads to the thermodynamic length:
\be 
\lambda = \left ( r_+ - KM \frac{1+\frac{a^2}{\ell^2}}{1-\frac{a^2}{\ell^2}} \right)
= \left ( r_+ - \frac{KM}{L}(2-L) \right)
\ee
Thus our black hole obeys a standard first law, 
\be
\delta M = T \delta S + V \delta P + \Omega \delta J + \Phi \delta Q
-2 \lambda \delta\mu
\ee
as well as a conventional Smarr relation:
\be
M
= 2TS + \Phi Q - 2 PV + 2 \Omega J
\ee

Once the black hole is accelerating, it is no longer possible to have a completely 
non-rotating frame at infinity due to the distortion of the boundary. The potential 
therefore has a more complicated adjustment involving not only a shift, but also
a decision on the appropriate frame to use at infinity. Computing the relevant 
parameters for the rotating accelerating black hole is underway.

\section*{Acknowledgements} 

We would like to thank Gary Gibbons, Rob Myers and Claude Warnick 
for discussions on the C-metric and conical deficits.
MA is supported by an STFC studentship.
RG is supported in part by STFC (Consolidated Grant ST/L000407/1),
in part by the Wolfson Foundation and Royal Society, and also by the
Perimeter Institute for Theoretical Physics. DK is also
supported in part by Perimeter, and by the NSERC.
Research at Perimeter Institute is supported by the Government of
Canada through the Department of Innovation, Science and Economic 
Development Canada and by the Province of Ontario through the
Ministry of Research, Innovation and Science.
RG would also like to thank the Aspen Center for Physics for hospitality.
Work at Aspen is supported in part by the National Science Foundation 
under Grant No.\ PHYS-1066293.

\providecommand{\href}[2]{#2}
\begingroup\raggedright

\end{document}